\documentclass[twocolumn,english,aps,pre,superscriptaddress,showpacs]{revtex4}
\usepackage[T1]{fontenc}
\usepackage[latin9]{inputenc}
\usepackage{amsmath}
\usepackage{graphicx}
\usepackage{amssymb}
\usepackage{nicefrac}
\usepackage{dsfont}
\usepackage[normalem]{ulem}
\usepackage[usenames]{color}

\usepackage{graphicx}
\usepackage{amsmath,float}

\makeatletter
\renewcommand \thetable {\@Roman\c@table}
\makeatother

\begin{document}
\title{Elastic response of filamentous networks with compliant crosslinks}

\author{A. Sharma}
\affiliation{Department of Physics and Astronomy, Vrije Universiteit, Amsterdam, The Netherlands}

\author{M. Sheinman}
\affiliation{Department of Physics and Astronomy, Vrije Universiteit, Amsterdam, The Netherlands}

\author{K. M. Heidemann}
\affiliation{Institute for Numerical and Applied Mathematics, G$\ddot{o}$ttingen University, Germany}

\author{F. C. MacKintosh $\footnote[1]{Author to whom correspondence should be addressed. Electronic mail: fcm@nat.vu.nl}$}
\affiliation{Department of Physics and Astronomy, Vrije Universiteit, Amsterdam, The Netherlands}

\date{\today}

\begin{abstract}
Experiments have shown that elasticity of disordered filamentous networks with compliant crosslinks is very different from networks with rigid crosslinks. Here, we model and analyze filamentous networks as a collection of randomly oriented rigid filaments connected to each other by flexible crosslinks that are modeled as worm-like chains. For relatively large extensions we allow for enthalpic stretching of crosslinks' backbones. We show that for sufficiently high crosslink density, the network linear elastic response is affine on the scale of the filaments' length. The nonlinear regime can become highly nonaffine and is characterized by a divergence of the elastic modulus at finite strain. In contrast to the prior predictions, we do not find an asymptotic regime in which the differential elastic modulus scales linearly with the stress, although an approximate linear dependence can be seen in a transition from entropic to enthalpic regimes. We discuss our results in light of the recent experiments.
\end{abstract}

\maketitle

The mechanical response of animal cells is largely determined by a visco-elastic matrix known as the \emph{cytoskeleton}, which is a network consisting of many different biopolymers together with various binding proteins that govern the organization and stability of these networks. One of the ways in which these networks differ from synthetic crosslinked polymer systems is the fact that many of the crosslinks are themselves highly compliant proteins, which can strongly affect the macroscopic network compliance. There have been many \textit{in vitro} studies of reconstituted networks with rigid crosslinks\cite{Janmey1991,Janmey1997,Pollard1998,Weitz2004,WeitzPRL2004,Janmey2005,Kroy2006,Koenderink2006,Fletcher2007,Janmey2007,Weitz2007,LiuKoenderinkPRL2007}. By comparison, relatively few recent experimental or theoretical studies have focused on networks with compliant crosslinks\cite{Bausch2006,Gardel2006PNAS,Gardel2006PRL,Levine2007,ChasePRL2008,Lubensky2007,Lee2009}.

A model experimental system of filamentous networks with compliant crosslinks is that of F-actin networks with the highly compliant and physiological crosslink \emph{filamin}. Experimental studies on such networks have demonstrated several striking elastic properties: These networks can have a linear modulus as low as 1 Pa, while able to withstand stresses as large as 100 Pa or more at strains of order 1 or less. They do so by stiffening dramatically by up to a factor of 1000 under applied shear stress. Moreover, in contrast to networks with noncompliant crosslinks, such networks can be subjected to relatively high strains $\gtrsim50\%$ without rupturing\cite{Gardel2006PRL,Weitz2007}. Both the linear and nonlinear elastic properties of actin-filamin gels appear to be dramatically affected by the flexible nature of the crosslinks, resulting in novel behavior as compared to actin-networks with noncompliant crosslinks, and to synthetic polymer gels. Similar composites of rigid filaments and compliant crosslinks can be found in other systems, such as stiff DNA nanotubes crosslinked by flexible DNA linkers\cite{Omar2012}, although much less is known about the nonlinear elastic properties of such systems.%

Recently, a model for the elastic behavior of such networks has been proposed\cite{ChasePRL2008,ChasePRE2009}. The model is based on the assumption that the soft stretching modes of the crosslinks govern the elasticity of a dense network of stiff polymers. Specifically, the filaments are assumed to be rod-like and much more rigid than the crosslinks. Moreover, these rods are assumed to be much longer than the contour length of the fully-stretched crosslinks. Both of these assumptions are reasonable for actin-filamin networks. In this model the nature of the deformation field in the network in response to an externally imposed strain is determined in the following way. If the network response to an externally imposed strain is to produce a uniform deformation field, that would imply stretching of the rods, which in the limit of infinite rigidity of the rods, is energetically prohibited. It follows, that the deformation field should be non-uniform on scales below the rod length. On the macroscopic scale the network deforms affinely and stretches the crosslinks on length scales smaller than the length of the rod.

In the linear elastic regime the crosslinks behave as Hookean springs. However, since the deformation field is non-uniform on the length scale smaller than the full rod contour length, crosslinks at the outermost ends of the rod experience more stretching than those near the center of the rod. At a critical strain the outer-most crosslinks reach their full extension and, consequently, stiffen dramatically. This model predicted an onset of stiffening, that depends only on the rod length $L$ and the fully-stretched crosslink contour length $l_0$. Also, the linear modulus was predicted to be proportional to the mean number of crosslinks and $L$. This model accounts for many important features of the measured elastic response of such networks in the linear elastic regime, including the dependence on length $L$ \cite{Kasza2010}. The model also predicted a linear scaling of nonlinear elastic stiffness on applied stress, in agreement with numerous observations from experiments on actin-filamin gels \cite{Gardel2006PNAS,Bausch2006,Kasza2009}.

Here we analyze numerically and analytically filamentous networks with compliant crosslinks. Our simulations allow us to test both our analytic results, as well as the predictions and assumptions of the previous studies in Refs.\ \cite{ChasePRL2008,ChasePRE2009}. Importantly, certain key assumptions made in the earlier mean-field model were not tested. Among the important aspects of this model was the puzzling prediction of a weaker divergence of the stress with strain than would be expected for the crosslinks alone, in spite of the assumption that the rods connecting the crosslinks are noncompliant. More precisely, the model predicted that the stress and elastic modulus only diverge at infinite strain despite a finite-strain divergence of a crosslinks' spring constant.

Our networks are composed of randomly oriented rods of length $L$ in 2D, which are linked together by highly flexible crosslinks. Each crosslink is modeled as a worm-like-chain (WLC)\cite{Smith1994,Siggia1995}. The crosslinks consist of two binding domains interconnected by a thermally fluctuating flexible polymer chain of length $l_0$, assumed to be larger than the chain persistence length $\ell_p$. The compliance of such a crosslink is entropic in nature. Indeed, atomic force microscope (AFM) measurements show that a compliant actin crosslink filamin can be accurately described as a WLC\cite{Yamazaki2001,Rief2003}. In our simulations, we do not take into account domain-unfolding of the crosslink proteins\cite{Gardel2006PNAS,Gardel2006PRL,Levine2007}. It has been recently shown that the macroscopic stress required for domain unfolding exceeds the typical limit of shear stress at which network failure occurs\cite{Kasza2009,Kasza2010}. In fact, unbinding of filamin occurs at forces much smaller than that needed for unfolding \cite{Lang2008}. Therefore we consider only the stiffening of the crosslinks originating from the WLC model. Additionally, since the forces become unphysically high as the crosslink stretches to its contour length, we perform simulations where we allow for enthalpic stretching of the crosslink backbone. We also consider the scenario where rods can undergo enthalpic stretching as well. %

As in Refs.~\cite{ChasePRL2008,ChasePRE2009}, we also study this model analytically, assuming that the network deforms affinely on scale larger than the length of the rods. Under this assumption the deformation field on scales below the rod length is obtained for a given orientation of the filament. We then calculate the free energy stored in the crosslinks for an externally imposed strain, from which we obtain elastic modulus and shear stress of the network. We refer to this approach as the affine approximation.

Our simulation results are in excellent agreement with the predictions of Refs.\ \cite{ChasePRL2008,ChasePRE2009} for the linear elastic regime. However, in the nonlinear stiffening regime, there are significant differences. In particular, we do not find the regime predicted by the model proposed in Refs.\ \cite{ChasePRL2008,ChasePRE2009} in which the differential modulus scales linearly with stress. We suggest that the linear dependence of nonlinear elastic stiffness on the applied stress observed in experiments is not a power-law regime. Instead, it corresponds to a crossover from stiffening to enthalpic regime. We show that in networks with crosslinks that cannot undergo enthalpic stretching, independent of the crosslink density, there is a unique power-law stiffening regime where differential modulus scales with an exponent $3/2$ with the shear stress. The transition from the linear elastic regime to the $3/2$ power-law scaling of modulus with stress is a crossover of which the width depends on the crosslink density. Moreover, there is a maximal strain that can be applied to a network of stiff filaments with WLC crosslink. We show that the linear elastic regime of dense networks can be fully understood by invoking affine deformation on the length scale of filaments. In the nonlinear regime, the assumption of affine deformation on scale of filament's length breaks down. This is particularly true of sparse  networks, which we also study. For all strains, the displacement field in a sparse network is nonaffine on the length scale of filament. We show that by a simple empirical correction to the crosslink density, the linear modulus of sparse as well as dense networks can be expressed by a single expression. %

The remainder of the article is organized as follows. First, in Sec.~\ref{Model} we briefly describe the simulation model and study the linear elastic regime of a densely crosslinked network. In Sec.~\ref{LinearRegime} an expression for the linear modulus is derived and compared with the simulation results. In Sec.~\ref{nonlinearregime} the model is analyzed in the nonlinear regime with and without enthalpic stretching of crosslinks and filaments. Finally, in Sec.~\ref{sparsenetworks} we focus on the sparse networks. We also consider the scenario with a torsional rigidity of a crosslink. We show that near the rigidity percolation, the linear modulus of the network is determined by both the torsion constant and the spring constant of the crosslinks in an anomalous coupled manner.


\begin{figure}[t]
\begin{center}
  \includegraphics[width=\columnwidth]{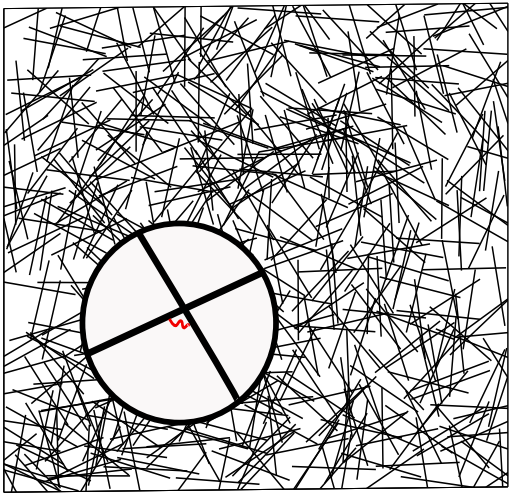}\\
  \caption{Schematic of mikado filamentous network. A crosslink is inserted at every intersection of any two filaments. A crosslink connecting two filaments is shown in red in the zoomed up intersection of two filaments.}\label{Networkschematic}
  \end{center}
\end{figure}

\begin{figure}[t]
\begin{center}
  \includegraphics[width=\columnwidth]{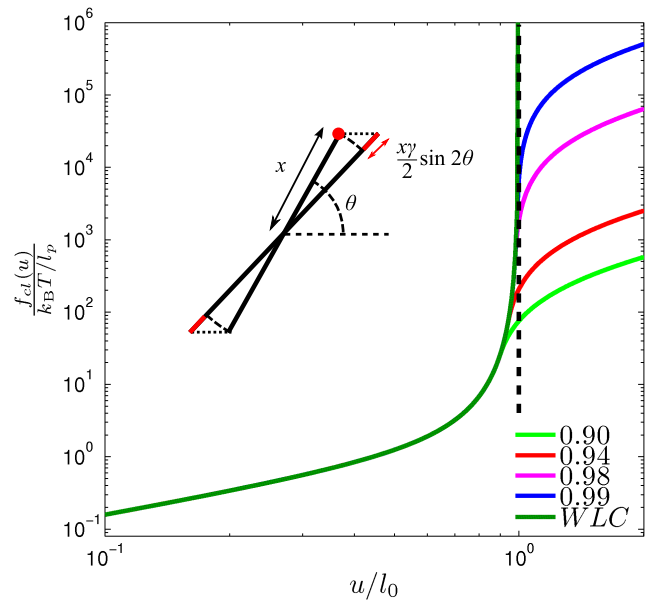}\\
  \caption{Force-extension curve of a worm like chain crosslink, including enthalpic stretching. Crosslink becomes a hookean spring when $u \gg l_{ee}$, with stiffness equal to the differential stiffness of the WLC curve at $u=l_{ee}$. Different curves correspond to different values of $l_{ee}/l_0$. Dashed line marks the divergence of force when a crosslink is stretched to it's contour length. Inset: In the rest frame of the filament, the deformation appears as pure rotation indicated by the dashed arc. Crosslink shown as red filled circle at distance $x$ from the center of filament stretches by $x\gamma \sin(2\theta)/2$.}\label{WLCforcecurve}
  \end{center}
\end{figure}

\section{Simulation Model}\label{Model}

We perform two-dimensional numerical simulations of disordered filamentous networks. The networks are modeled as collections of $N$ filaments/rods each of length $L$ distributed randomly within an area $A = W \times W$. Unless otherwise is specified, the system size is fixed to $W=10$. The initial orientation as well as the initial location of center of mass of the filaments is uniformly randomly distributed. A schematic of the filamentous network is shown in Fig.~\ref{Networkschematic}. Unless otherwise is specified, wherever two filaments intersect we insert a crosslink with rest length $0$. The force-extension curve of a flexible crosslink assuming that the contour length is larger than the persistence length is described by the formula\cite{Siggia1995,Lang2008}
\begin{equation}\label{WLC}
    f_{\rm cl}(u) = \frac{k_{\rm B}T}{l_p}\left(\frac{1}{4\left(1-\frac{u}{l_0}\right)^2}-\frac{1}{4}+\frac{u}{l_0}\right),
\end{equation}
where $k_{\rm B}T$ is the thermal energy, $l_p$ is the persistence length, $u$ is the extension, and $l_0$ is the contour length of the crosslink. The force-extension curve shown in Fig.~\ref{WLCforcecurve} demonstrates the rapid stiffening of the crosslink with diverging force as the extension approaches its contour length. We can include enthalpic stretching of the crosslink by treating the crosslink as a hookean spring beyond a certain extension $l_{\rm ee} < l_0$. In Fig.~\ref{WLCforcecurve} we show the force-extension curves for different values of $l_{\rm ee}/l_0$. For $u\ll l_0$, $f_{\rm cl}(u)=\frac{3}{2}\frac{k_{\rm B}T}{l_pl_0}u$ implying that the crosslink is a Hookean spring in the small extension limit with spring constant $k_{\rm cl}$ given by $\frac{3}{2}\frac{k_{\rm B}T}{l_pl_0}$. This defines an effective potential that we shall call the free energy.

For sparse networks, we also include the energy associated with change in the relative orientation of two filaments at a crosslink. We assume that this change in energy corresponds to torsion of the crosslink. The corresponding free energy is assumed to be quadratic as $F_{\tau} = 0.5k_{\tau}\left(\theta - \theta_0\right)^2$, where $k_{\tau}$ is the torsion constant of the crosslink and $(\theta-\theta_0)$ is the change in the relative angle between the two filaments connected by the crosslink. The average number of crosslinks per filament is determined by the the line density $\rho=NL/A$. If $\rho$ is sufficiently high, the average number of crosslinks per filament is given by $n=2\rho L/\pi$ \cite{HeadPRE2003}. In the case of infinite stiffness of the filaments, three degrees of freedom are associated with each filament. If the filaments can undergo enthalpic stretching, two additional degrees of freedom are associated at the location of each crosslink on the filament. The shear strain, $ \gamma $, is applied using Lees-Edwards periodic boundary conditions. The network response to any externally imposed strain is calculated by applying first an affine deformation to the filaments' centers of mass and then letting the network relax to it's mechanically equilibrium state using conjugate gradient algorithm \cite{CRecipes}. The free energy of the network in this state, $ F $, is then used to extract the modulus as $K = \frac{1}{A}\frac{\partial^2 F}{\partial\gamma^2}$ and shear stress as $\sigma = \frac{1}{A}\frac{\partial F}{\partial \gamma}$.

\section{Linear Regime}\label{LinearRegime}
In the linear elastic regime, the strain is low enough such that no crosslink reaches its full extension. We derive here an expression for the linear modulus $G_0$ assuming that the network deforms affinely on length scale of the filament. By imposing affine deformation on length scale of the filament, the deformation field, $u$, on sub-filament level is fully determined. Since a crosslink is connected to a pair of filaments, one can obtain an exact expression for the amount of stretching in a crosslink for a pair of filaments. In the small strain approximation, the contribution to the change in the stretching energy in a crosslink due to the change in the relative orientation of the two filaments averages out to zero. In the following derivation we ignore this contribution. Consider the response of crosslinks on a single filament to a small strain $\gamma$. In the rest frame of the filament a crosslink at at distance $x$ from the center is stretched by $u(x,\gamma)\propto x\gamma$. The total free energy of the network can be obtained by summing over all crosslinks as
\begin{equation}
F=Nk_{\rm cl}\int_0^{L/2}\frac{n}{L}u^2(x,\gamma) dx .
\label{Elin}
\end{equation}

In 2D, for a filament oriented at an angle $\theta$, $u(x,\gamma) = x\gamma \sin(2\theta)/2$ in the small strain approximation, as shown in the inset of Fig.~\ref{WLCforcecurve}. Averaging free energy over the uniformly distributed orientation of the filament, we extract the linear modulus as $G_0 = \frac{2F}{A\gamma^2}$, which yields
\begin{equation}
G_0= \frac{1}{96}\rho k_{\rm cl}nL.
\label{G0}
\end{equation}

\begin{figure}[t]
\begin{center}
  \includegraphics[width=\columnwidth]{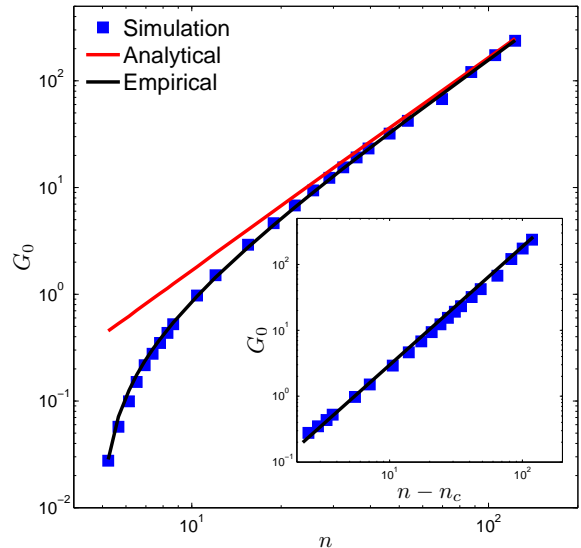}\\
  \caption{Linear modulus as function of the number of cross-links per filament. Red solid line corresponds to the analytical prediction of Eqn.~\eqref{G0}. The black line corresponds to the empirical correction to Eqn.~\eqref{G0} in which $n$ is replaced by $n-n_c$, where $n_c=4.93$. Inset shows $G_0$ obtained from simulations and the empirical fit plotted with respect to the reduced cross-link density.}\label{LinModSim}
  \end{center}
\end{figure}

The numerical pre-factor depends on the dimensionality. In 1D it is $1/8$ and in 3D it is $1/192$\cite{ChasePRL2008,ChasePRE2009}. In Fig.~\ref{LinModSim} we plot the linear modulus obtained from simulations versus the number of cross-links per filament. It is clear from the figure that $G_0$ from Eqn.~\eqref{G0} is in agreement with the simulation results for $n \gg 10$. For $n \ll 10$ linear modulus obtained in simulations differs substantially from the analytical prediction. This is expected because for low densities, the assumption of uniformly distributed crosslinks is no longer valid. Moreover, $n \ll 10$ corresponds to line densities for which the deformation of the network is not expected to be affine on the length scale of the filaments. In fact the network is not even rigid for $n \leq n_c$ where $n_c\approx4.93$. $n_c$ is the critical number of cross-links per filament at the rigidity percolation transition\cite{HeadPRE2003}. If $n$ in Eqn.~\eqref{G0} is replaced by $(n-n_c)$ we obtain a very good agreement with the analytical prediction over the whole range of $n$, as shown in Fig.~\ref{LinModSim}. Although we have introduced this correction as purely empirical, it can be justified by analogy with mean-field theory of rigidity percolation. In the mean field description of the rigidity near percolation point, all macroscopic quantities of interest, such as modulus and stress, scale linearly with the distance from the rigidity percolation point\cite{Thorpe1985}. The numerical and analytical results in the linear elastic regime, presented so far, agree with the previous analytical modeling\cite{ChasePRL2008,ChasePRE2009}. However, as we show in the next section, in the nonlinear regime this is not the case.

\section{Non Linear Regime}\label{nonlinearregime}

\begin{figure}[t]
\begin{center}
  \includegraphics[width=\columnwidth]{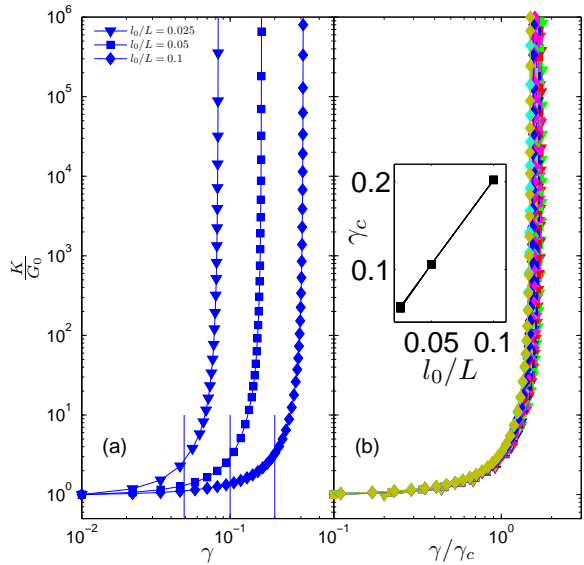}\\
  \caption{(a) Differential modulus in units of linear modulus $G_0$ for $n=70$. The small vertical lines mark the critical strain $\gamma_c$. (b)Same as (a) but now strain is expressed in units of $\gamma_c$. Different colors correspond to different values of $n$. The inset shows the critical strain $\gamma_c$ for different values of $l_0$.}\label{NlinModSim}
  \end{center}
\end{figure}

We now consider strain induced stiffening of filamentous networks with compliant cross-links. In these networks the onset of stiffening corresponds to the strain at which cross-links, lying at the edge of a filament, approach their contour length. Under the assumption that network deformation is affine on the length scale of the filament, a cross-link at the edge of a filament stretches by an amount $dl \simeq \gamma L/2$. When $dl \simeq l_0$, the cross-link will stiffen dramatically. It follows, that the strain $\gamma_c \simeq 2l_0/L$ marks the onset of nonlinear stiffening of the network\cite{ChasePRL2008,ChasePRE2009}. In Fig.~\ref{NlinModSim} we plot the differential shear modulus, $K$, in units of the linear modulus $G_0$, as a function of strain for different values of $l_0$. In simulations we define $\gamma_c$ as the strain at which $K/G_0 \sim 3$. We plot $\gamma_c$ in the inset of Fig.~\ref{NlinModSim}(b). Each point in the curve corresponds to five different values of $n$, ranging from $n=30$ to $n=80$.  These points are indistinguishable because $\gamma_c$ is determined solely by the ratio $l_0/L$. It is clear from the figure, that $\gamma_c$ obtained from simulations scales linearly with $l_0/L$ and is independent of the cross-link density. %

An important feature of the non linear stiffening is that the differential modulus diverges at a particular strain $\gamma_d \sim \pi l_0/L$. In other words, there is a maximal strain that can be applied to networks with cross-links with an unstretchable backbone. The fact that the strain is bounded is in contrast with the predictions of the previous work Refs.\ \cite{ChasePRL2008,ChasePRE2009}. There, the differential modulus is finite for any strain. This result follows from the assumption that beyond critical strain the cross-links, that have already stretched to their full extension, become infinitely rigid and any strain beyond critical strain leads to stretching of the medium. In other words, beyond critical strain the medium starts stretching and the medium's elastic response, although highly nonlinear, is not divergent for any finite strain. This assumption about the medium would be valid, provided the medium existed independent of the constituting filaments, such that each filament is connected to the medium rather than to each other. Our simulations show that the weak divergence of modulus and strain is a non-physical prediction of the model. If the filaments are infinitely rigid and crosslinks have finite compliance, both modulus and stress will diverge at a finite strain.

In the previous work, the linear scaling of the differential modulus with the stress was a consequence of the stretching of the medium beyond the critical strain. It is clear that such a regime is not observed in our simulations of networks with WLC crosslinks. However, experiments have shown that the differential modulus depends linearly on the applied shear over a significant range of stress. Moreover, such networks can be subjected to strains much larger than the critical strains. We will show below that these experimental observations can be understood by allowing for enthalpic stretching of the crosslinks beyond their contour length.
\begin{figure}[t]
\begin{center}
  \includegraphics[width=\columnwidth]{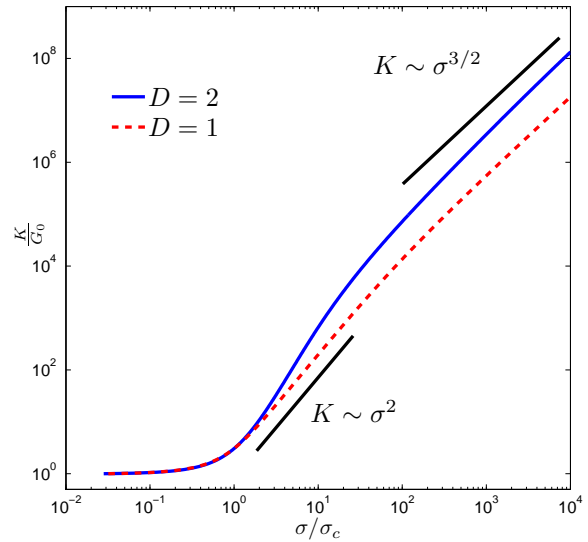}\\
  \caption{Affine calculation: Differential modulus in units of linear modulus $G_0$ for $l_0/L=0.1$ for one and two dimensional filamentous  networks with compliant cross-links.}\label{affinecalculation}
  \end{center}
\end{figure}

\subsection{Deviation from the affine approximation}

Under the affine approximation the exact form of divergence of differential modulus with strain can be determined in a straightforward manner. Here we consider the case of extremely high, but finite cross-link density. In 2D the magnitude of the displacement field depends on the orientation of the filament and can be written as

\begin{equation}
u(x,\gamma)=
x\left(\sqrt{1+\gamma^2 \sin^2\theta + \gamma \sin2\theta}- 1\right),
\end{equation}
where $x$ is the distance of the cross-link from the center of the filament.
We use the expression of Eqn.~\eqref{WLC} to calculate the energy of cross-links on a single filament as
\begin{equation}\label{Enlin}
    F(\gamma) = \frac{2n}{L}\int_0^{L/2}\int_0^{u(x,\gamma)}f_{\rm cl}(u)du dx.
\end{equation}
We obtain energy by numerically integrating Eqn.~\eqref{Enlin} and averaging over all orientations of the filament. The results for differential modulus vs stress are plotted in Fig.~\ref{affinecalculation}. As can be seen there, the nonlinear stiffening is characterized by a broad crossover followed by power-law scaling of differential modulus with shear stress with an exponent $3/2$.  However, the transition to the $3/2$ regime occurs at very high modulus and, therefore, may not be relevant for experiments. Also, in Fig.~\ref{affinecalculation} the results for the 1D case are shown. The differential modulus scales quadratically with the shear stress before rolling onto the $3/2$ power-law regime. The quadratic scaling of $K$ with $\sigma$  follows directly from a one dimensional displacement field $u(x,\gamma)\sim x\gamma$. Using this one dimensional displacement field in Eqn.~\eqref{Enlin}, we obtain that near the strain $\gamma_d$, $E\sim -\ln\left(1-\gamma/\gamma_d\right)$. It follows that in the vicinity of $\gamma_d$, stress $\sigma \sim \left(1-\gamma/\gamma_d\right)^{-1}$ and differential modulus $K \sim \sigma^2$. However, the width of the quadratic regime depends on the cross-link density and only for the unphysical limiting case of infinite cross-link density, this regime persists over the entire stress range. For any finite cross-link density the differential modulus will scale as $K\sim\sigma^{3/2}$ for high stresses as shown in Fig.~\ref{affinecalculation}.

\begin{figure}[t]
\begin{center}
  \includegraphics[width=\columnwidth]{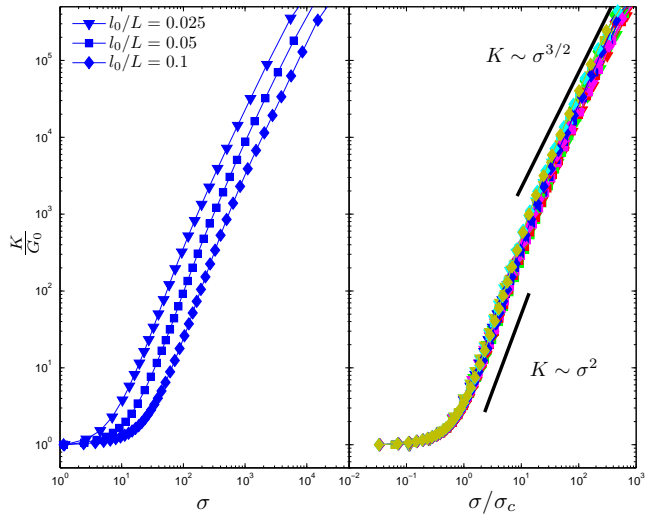}\\
  \caption{(a) Differential modulus in units of linear modulus $G_0$ for $n=70$. (b)Same as (a) but now stress is in units of $\sigma_c$. Different colors correspond to different values of $n$. The modulus scales approximately as $K\sim \sigma^2$ before it enters the power-law regime $K\sim \sigma^{3/2}$.}\label{ModvsStress}
  \end{center}
\end{figure}

\begin{figure}[t]
\begin{center}
  \includegraphics[width=\columnwidth]{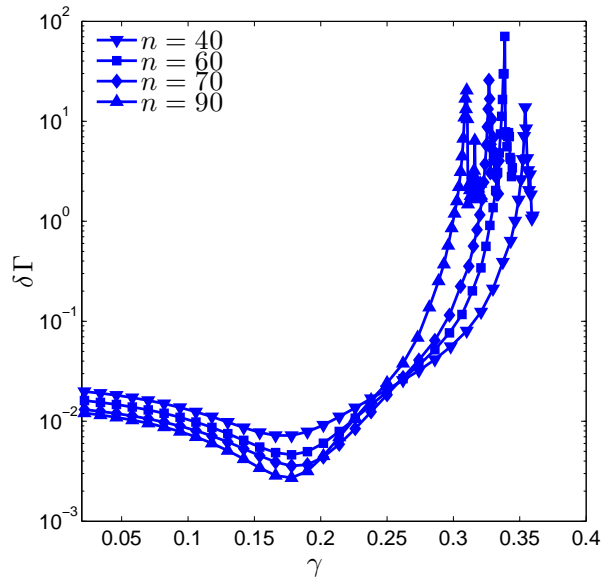}\\
  \caption{Differential non-affinity as a function of strain for different cross-link densities. The maxima in the differential non-affinity shifts towards left with increasing cross-link density.}\label{diffNA}
  \end{center}
\end{figure}

It is clear from Fig.~\ref{affinecalculation} that the affine elastic response of a 2D network exhibits an initially steeper than quadratic dependence on the shear stress in the crossover.
In Fig.~\ref{ModvsStress} we plot the differential shear modulus $K$ obtained from simulations, in units of linear modulus $G_0$, as a function of shear stress for different values of $l_0$. We define critical stress $\sigma_c = \sigma(\gamma_c)$. Stiffness curves of $K/G_0$ vs. $\sigma/\sigma_c$, corresponding to different values of cross-link density, $n$, collapse, as shown in Fig.~\ref{ModvsStress}(b). As can be seen in the figure, for relatively high $\sigma$ values $K\sim\sigma^{3/2}$. However, the elastic response of the network in the crossover is softer than the one expected from the two-dimensional affine approximation but stiffer than that from the 1D approximation. The quadratic dependence, as shown in Fig.~\ref{ModvsStress}(b), is an approximate dependence and corresponds to a crossover. In fact, with increasing cross-link density, the crossover is characterized by a steeper than quadratic dependence of modulus on stress. Clearly the affine approximation is invalid in the nonlinear regime. The degree of nonaffine deformations in the network can be quantified as:
\begin{equation}
\delta \varGamma(\gamma) = \frac{\langle dr_{\rm aff} - dr\rangle^2}{N d\gamma^2},
\end{equation}
where $\delta \varGamma$ is referred to as the differential non-affinity, $dr_{\rm aff}$ is the affine displacement of center of mass of the filament to an incremental strain $d\gamma$, $dr$ is the actual displacement of the center of mass of the filament, and $N$ is the total number of filaments in the network. $\delta \varGamma$ is a measure of the nonaffine displacement of the filaments to an incremental strain. Affine approximation predicts that $\delta \varGamma = 0$. In Fig.~\ref{diffNA}, we plot $\delta \varGamma$ in the network as a function of the applied strain. In the linear elastic regime $\delta \varGamma$ is small consistent with the assumption of affine displacements. As $\gamma$ approaches $\gamma_d$, $\delta \varGamma$ increases rapidly implying that the affine approximation on scale of filament's length is no longer valid.

\begin{figure}[t]
\begin{center}
  \includegraphics[width=\columnwidth]{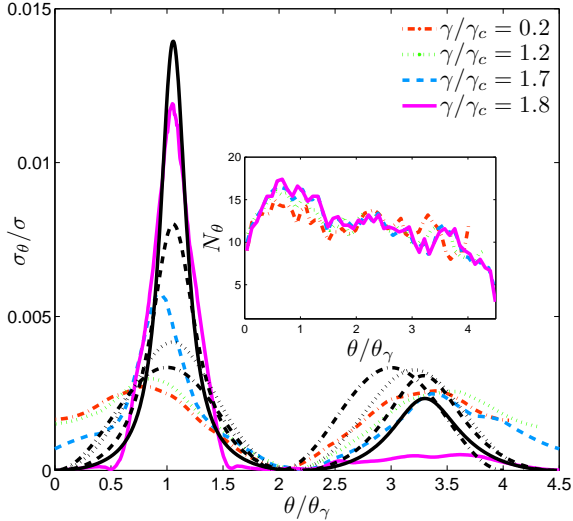}\\
  \caption{Average stress per filament as a function of the orientation angle of the filaments. The orientation angle is expressed in units of the angle $\theta_{\gamma} = \tan^{-1}\left(1/(1+\gamma)\right)$ corresponding to the strain. The peak corresponds to the filaments oriented along the direction of the applied strain. The secondary smaller peak corresponds to the filaments oriented against the direction of the applied strain. Black lines corresponds to the affine approximation. The inset shows the angular distribution of filaments. Besides a small increase in the number of filaments along the direction of strain, the distribution is practically uniform.}\label{selectivepaths}
  \end{center}
\end{figure}

As mentioned above, the elastic response of the network is softer than the one expected from the two-dimensional affine approximation.
In Fig.~\ref{selectivepaths} we plot the stress per filament as a function of the orientation of the filament obtained from simulations and the affine approximation. It is clear from the figure that filaments oriented along and against the direction of the applied strain bear most of the stress in the network. For low strains, corresponding to the linear elastic regime, the contribution coming from paths along and against the direction of strain are practically equal. With increasing strain, the relative contribution of filaments oriented against the direction of the strain decreases evident as the decreasing height of the secondary peak in Fig.~\ref{selectivepaths}. We also show in Fig.~\ref{selectivepaths} the angular distribution of the filaments for different strains. That the distribution of filaments is practically uniform implies that there is insignificant geometrical alignment of filaments along the direction of strain. It is clear from the figure that for a given strain the affine approximation overestimates the stress. The relative contribution to the total shear stress from the filaments oriented against the direction of strain decreases with increasing strain but remains substantially higher than those obtained from simulations. It follows that with increasing strain paths oriented along the direction of strain determine the elastic response of the network.

\begin{figure}[t]
\begin{center}
  \includegraphics[width=0.95\columnwidth]{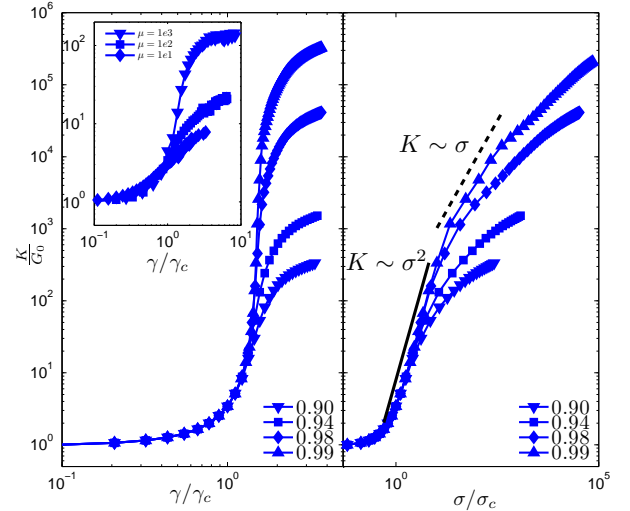}\\
  \caption{(a) Differential modulus in units of linear modulus $G_0$ for $n=70$ versus $\gamma/\gamma_c$ for different values of $l_{ee}/l_0$. The modulus converges to a constant at high values of strain. The inset shows the differential modulus when filaments are compliant. (b)Differential modulus plotted with respect to stress in units of critical stress. Black solid line indicates the approximate $K\sim \sigma^2$. The dashed line with unit slope is guide to eye to show that the crossover regime approximately follows $K\sim \sigma$  }\label{enthalpic}
  \end{center}
\end{figure}

\subsection{Enthalpic stretching}
Experimentally it has been shown that filamentous networks with compliant cross-links can be subjected to very high strains $\gtrsim50\%$\cite{Gardel2006PRL,Weitz2007}. In our simulations the network can be subjected to any strain by allowing for enthalpic stretching of the cross-links. In that case we expect that the large strain limit of the differential modulus will correspond to the enthalpic stiffness of the cross-links. In Fig.~\ref{enthalpic} we plot the differential shear modulus $K$ in units of linear modulus $G_0$ as a function of strain in (a) and with shear stress in (b) for different values of $l_{ee}/l_0$. As can be seen in Fig.~\ref{enthalpic}(a), on adding enthalpic stretching of the cross-links, the differential modulus does not show any divergence. Instead, the modulus converges to a constant for higher strains. The high-strain limit of the differential modulus is governed by the cross-link density and the enthalpic stiffness of the cross-links. An interesting feature is evident in the $K$ vs. $\sigma$ plot in Fig.~\ref{enthalpic}(b). In the initial stiffening regime, the modulus increases steeply with the stress. For higher stresses, there is a crossover in which the modulus approaches a constant determined by the enthalpic stretching modulus of the cross-link backbone. In this crossover one could interpret the dependence of modulus as linear in stress, as shown in the figure. In simulations the stress can increase without bounds but in experiments the network fails beyond a certain stress. We think that it is this crossover in experiments where an approximate linear dependence of modulus on stress has been inferred.

We also performed simulations by considering that, in addition to the cross-links, filaments can undergo stretching as well, characterized by stretching modulus $\mu$. As long as $k_{cl} \ll \mu/\tilde{l_c}$, where $k_{cl}$ is the differential stiffness of the cross-link and $\tilde{l_c}$ is the average cross-link distance, the stiffness of the network is determined entirely by the cross-links. At high strains cross-links become stiff, such that further deformation of the network leads to stretching of filaments. As expected, the differential modulus converges to a value governed by $\mu/\tilde{l_c}$, as shown in the inset of Fig.~\ref{enthalpic}(a). Whether this regime, where filaments undergo stretching in a network, is accessible or not depends on the cross-link density and the binding strength of the cross-links. If the cross-link density is very high or multiple cross-links are attached in very close proximity to each other, the local stiffness of filaments could fall below that of the cross-links. We emphasize that our simulations and calculations establish that there is no unique regime with linear scaling of modulus with stress.

\section{Sparse Networks}\label{sparsenetworks}

\begin{figure}[t]
\begin{center}
  \includegraphics[width=\columnwidth]{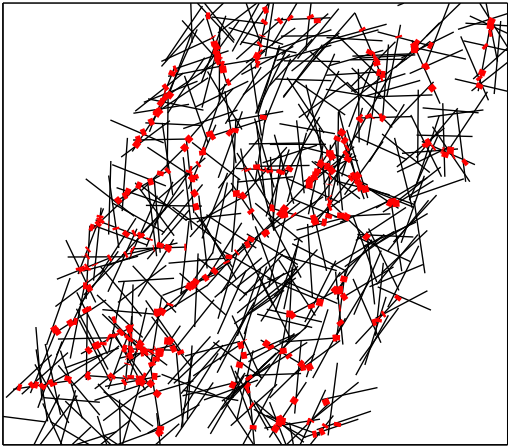}\\
  \caption{Network configuration at $\gamma = 2.5\gamma_c$ for $n=10$ and $l_0/L = 0.1$. Red lines indicate the set of stretched cross-links that sustain 90\% of the total stress in the network.}\label{sparsenetwork}
  \end{center}
\end{figure}

\begin{figure}[t]
\begin{center}
  \includegraphics[width=\columnwidth]{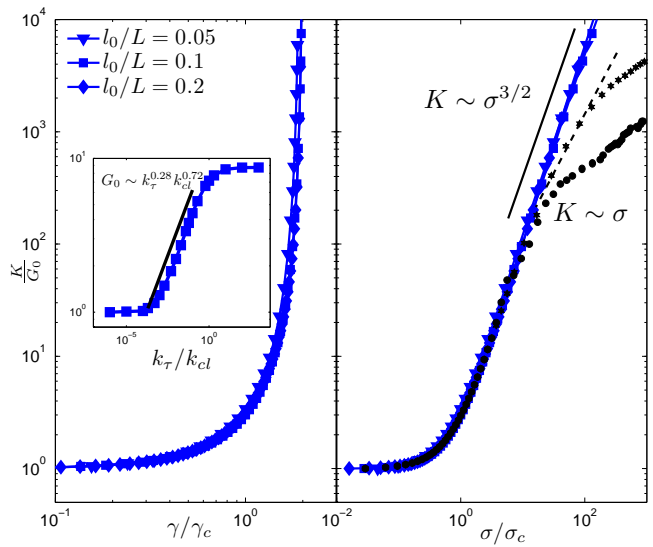}\\
  \caption{(a) Differential modulus in units of linear modulus $G_0$ for $n=10$ versus $\gamma/\gamma_c$.The inset shows the linear modulus $G_0$ at the rigidity percolation point. If rotation of filaments around cross-link costs energy, $G_0$ depends anomalously on torsion constant $k_{\tau}$ and $k_{cl}$. (b)Differential modulus plotted with respect to stress. Black solid line depicts $K\sim \sigma^{3/2}$ whereas the dashed line indicates the approximate $K\sim \sigma$. Filled circles ($l_{ee}/l_0= 0.92$) and stars ($l_{ee}/l_0= 0.96$) correspond to enthalpic stretching of the crosslinks.}\label{sparseNlin}
  \end{center}
\end{figure}

In networks with low cross-link density the affine approximation is invalid for any applied strain. The linear shear modulus $G_0$ is smaller than that obtained under the affine approximation, as shown in Fig.~\ref{LinModSim}. In Fig.~\ref{sparseNlin} we plot the differential modulus as a function of strain and stress. Similar to the high cross-link density networks, the differential shear modulus diverges at a certain strain $\gamma_d$ as $K~\sim \left(1-\gamma/\gamma_d\right)^{-3}$.  This results in a stiffening regime of the form $K\sim \sigma^{3/2}$, which extends till infinity. Same scaling of differential modulus with stress is observed in networks of semiflexible filaments with fixed cross-links. Interestingly, in networks of semi-flexible filaments, it is the affine deformation on length scale above the distance between the cross-links that results in this nonlinear scaling. We also plot results for the differential modulus as a function of stress allowing for enthalpic stretching of the crosslink, as shown in Fig.~\ref{sparseNlin}(b). Similar to the densely crosslinked networks, enthalpic stretching of the crosslinks' backbone results in a crossover in which the differential modulus depends approximately linearly on the stress. Moreover, as shown in Fig.~\ref{sparseNlin}(b), consistent with the experimental observations, the modulus shows a stiffening by factor of $\sim 1000$.

We suggest that sparse networks may be more relevant to the experiments as one does not expect the 3D filamentous networks to be densely crosslinked. In contrast to 2D networks, where an intersection between two filaments is unambiguously defined, in 3D random networks one needs a maximum distance criterion to consider two filaments as intersecting. Two filaments, such that the shortest distance between them is smaller than or equal to the distance set by the intersection criterion are considered intersecting. However, such procedure may result in initially pre-stressed networks. High crosslink density can be achieved in 3D networks by considering the filaments to bend on long length scales.

Deep into the stiffening regime, most of the stress in the network is carried by a small fraction of filaments. In Fig.~\ref{sparsenetwork} we show the network configuration highlighting the set of cross-links which carry about 90\% of stress in the network. In this configuration only 10\% of the filaments are carrying 90\% of the stress in the network. As was shown above, filaments that are oriented along the direction of strain sustain most of the stress.

In networks with fixed cross-links the modulus is determined by the stretching, $\mu$, and bending, $\kappa$, rigidity of the filaments. In lattice based models of the network, the linear modulus scales as $G_0 \sim \sqrt{\mu \kappa}$ at the central force isostatic point\cite{chase2011}. In the random filamentous networks, we study in this work, the modulus scales linearly with the stiffness of the cross-links in the linear elastic regime. In networks with compliant cross-links relative rotation of filaments around a cross-link will induce torsion in the cross-link. Considering that the torsion constant is given by $k_{\tau}$, we expect that the linear modulus $G_0$, at the rigidity percolation point, is determined by $k_{cl}$ and $k_{\tau}$ together in an anomalous fashion. Indeed, we obtain that near the rigidity percolation point $G_0 \sim k_{cl}^{\alpha_{\tau}}k_{\tau}^{1-\alpha_{\tau}}$, where the critical exponent $\alpha_{\tau}= 0.72\pm0.05$, as shown in the inset of Fig.~\ref{sparseNlin}. For low and high values of $k_{\tau}$, $G_0$ is independent of the torsion. In the intermediate range of $k_{\tau}$ network deformation couples to both torsion and stretching of cross-links, giving rise to the anomalous regime.  A recent theoretical study on networks of soft filaments crosslinked with flexible angle-constraining crosslinks  also reported a similar critical anomalous regime\cite{Das2012}. Due to the angle-constraining crosslinks, changing the relative orientation of the two bonds connected at the crosslink costs energy. The bond-bending scenario in the composite angle-constraining networks is analogous to the torsion in the crosslink in our filamentous networks connected with compliant crosslinks. However, unlike our sparse networks, the critical exponent obtained in the composite angle-constraining networks is $0.5$, consistent with the mean-field predictions\cite{Das2012}.
\section{Discussion and Conclusion}
In this article we studied random filamentous networks, linked together with compliant cross-links. We modeled the cross-links as worm like chains. We studied such networks, both analytically and numerically, by performing two-dimensional simulations. Our simulations show that for sufficiently high cross-linking density, network deformation in the linear elastic regime can be well approximated as affine on the scale of filament's length. As a consequence, we find good agreement with Refs.\ \cite{ChasePRL2008,ChasePRE2009}, e.g., in the fact that the linear shear modulus scales linearly with the length of the filament. Moreover, the onset of stiffening with increasing strain is independent of the cross-link density and is determined solely by ratio of contour length of cross-link to the length of the filament. We derived an expression for the linear modulus of two-dimensional filamentous networks. Numerical simulations in the linear elastic regime are in an excellent agreement with this analytical calculation as well as Refs.\ \cite{ChasePRL2008,ChasePRE2009}.

We also tested the predictions of Refs.\ \cite{ChasePRL2008,ChasePRE2009} in the nonlinear regime. In the effective medium model proposed in Refs.\ \cite{ChasePRL2008,ChasePRE2009}, the network can be subjected to any strain. This is because of the assumption that, beyond the critical strain, it is the medium that undergoes stretching, whereas the crosslinks, already stretched to their contour length, become infinitely rigid. We showed that this assumption of the medium undergoing stretching beyond the critical strain is not valid in filamentous networks with WLC crosslinks. In fact independent of the crosslink density, the stress and modulus of a filamentous network with WLC crosslinks diverges at a finite strain. Moreover, unlike the predictions of the effective medium model of linear scaling of differential modulus with the shear stress, no such regime is observed in the simulations in the appropriate parameter range. We conclude that the effective medium model is not valid in the stiffening regime of such networks.

We suggest that the effective medium model proposed in Refs.\ \cite{ChasePRL2008,ChasePRE2009} could account for the stiffening behavior of composite networks composed of rigid filaments, crosslinks and an embedding medium. Such composite networks can be described as random filaments connected to the underlying medium by crosslinks. In this scenario, the medium exists independent of the constituting filaments, which may make the mean-field approximation in Refs.\ \cite{ChasePRL2008,ChasePRE2009} a good one. For the networks we study in our simulations, however, it appears that the mean-field approximation is not valid, at least for the nonlinear response, for which our results differ substantially from the predictions of the mean-field or effective medium approach.

In the stiffening regime the assumption of affine deformation on the scale beyond filaments' length is no longer valid. Therefore, the elastic response of the network is softer than that expected from the affine approximation. For high cross-link densities, the initial part of the stiffening regime is characterized by a steep increase in the differential modulus with the shear stress. For high stresses the modulus exhibits $3/2$ power-law scaling with the shear stress. This $3/2$ power-law regime extends till infinity. On the other hand, for low cross-link densities, the entire stiffening regime is characterized by the $3/2$ power-law regime. In fact, for any finite cross-link density, the only persistent regime is that corresponding to the $3/2$ scaling of differential modulus with the shear stress. This particular form of scaling is well known in networks of semi-flexible chains with fixed cross-links under affine deformations. Interestingly, in networks with compliant cross-links, this scaling corresponds to nonaffine deformations on length scale of filaments and not to stiffening of individual fibers. The network response is governed by a relatively small set of filaments oriented along the direction of the strain. Such sparse networks develop highly stressed pathways in the network with increasing strain. Independently of the cross-link density, the differential modulus diverges at finite strain.

Enthalpic stretching of cross-links beyond a certain length implies that networks can be subjected to large strains. The same can be achieved by considering the filaments to be compliant. We considered both scenarios in our simulations. When cross-links can undergo enthalpic stretching the differential modulus of the network with high crosslink density exhibits an approximately quadratic dependence on the applied shear in the initial stiffening regime before rolling on to a regime where the modulus converges to a constant at high stresses. In sparse networks, in the stiffening regime, the differential modulus exhibits the $3/2$ power-law regime followed by crossover to a regime where the modulus converges to a constant at high stresses. Experimentally one might not observe the saturation of modulus as network failure may occur by unbinding of the cross-links. Over a relatively significant range of the crossover region, although not a power-law regime, differential modulus appears to depend linearly on the stress. It is likely that this crossover region has been interpreted as a regime with linear scaling of modulus with stress.

The specific choice of WLC approximation for the force-extension curve of compliant actin crosslinks is motivated by the AFM measurements on filamin \cite{Rief2003,Yamazaki2001}. However, the presented approach can be generalized to other systems with rigid filaments connected by compliant crosslinks. For instance, networks of stiff DNA nanotubes crosslinked by single-stranded flexible DNA\cite{Omar2012}. In this case the WLC approximation has to be replaced by freely-jointed-chain approximation\cite{Smith1996}. Under the freely-jointed-chain approximation the high stress dependence of the elastic modulus is quadratic in stress, in contrast to the $3/2$ power-law scaling for the WLC, considered here.

\section*{Acknowledgments}
We thank C.P. Broedersz, G. Koenderink, M. Wardetzky, C.F. Schmidt, M. Prabhune and F. Rehfeldt for helpful discussions. This work was funded in part by FOM/NWO and NanoNextNL of the Government of the Netherlands and 130 partners. Support by the colaborative research center SFB 755 (Nanoscale photonic imaging) is acknowledged.


\end{document}